\documentclass[12pt]{article}
\usepackage{graphicx}
\begin{document}
\begin{center}
{\Large \bf  Production of the  $\pi^{-}, \pi^{0}$  
in  the reaction $d(e,pp)e' \pi^{-} \pi^{0}$ \footnote{
This work was supported by Russian Foundation for Basic 
Research   
N 98-02-17993, 
N 98-02-17949 and by grant 
N 96-0424 
from INTAS.}. } \\ 	[5mm]
\end{center}
\begin{center}
\large{
{A.Yu.~Loginov$^a$, D.M.~Nikolenko$^b$, A.V.~Osipov$^a$, I.A.~Rachek$^b$,
A.A.~Sidorov$^a$,\\
V.N.~Stibunov$^a$, D.K.~Toporkov$^b$}\\ 
{\small \em $^a$Nuclear Physics Institute at Tomsk Polytechnic University,
Tomsk, Russia\\
$^b$Budker Institute of Nuclear Physics, Novosibirsk, Russia}
}
\\[6mm]
\end{center}
\begin{center}
\large
{Talk presented at the 15th International Conference\\
 on Particles and Nuclei,
 10-16 June, 1999,\\ Uppsala University, Uppsala, Sweden.}
\\[5mm]
\end{center}

\begin{abstract}
{
Recently, the differential yield and target asymmetry of the reaction\\ 
$ d(e,pp)e' \pi^{-} $ on the tensor polarized deuterium target have 
been presented in \cite{sid,few}, where  the experimental procedure and 
basic assumption were described. A precise treatment of this reaction 
needs to investigate the reaction channels with an additional neutral 
pion production as a  background to the leading reaction of a  single 
$\pi^{-}$ production.

We considered for this background process the pole mechanism and the 
two body mechanism. In the latter the $\Delta$-isobar and the pion 
are produced on one nucleon followed by the absorption of the pion 
on the second nucleon producing another $\Delta$-isobar. The final 
proton-pion systems come from the decay of the $\Delta^{+}$ and 
$\Delta^{0}$. 

In this work we estimated the contribution of the reaction with 
additional $\pi^{0}$  production into whole yield for the different 
polarization states of the deuteron.
}

\end{abstract}

In a study of the $d(e,pp)X$ - reaction at electron energy $2 GeV$, 
the momenta of both protons from 0.3 to 0.7 $GeV/c$ and proton 
emission angles $64^{\circ} - 83^{\circ}$ approximately 900 proton 
proton coincidences were analyzed \cite{sid,few}. The cross section of 
the process initiated by electrons was expressed in terms of 
the cross section of a reaction induced by virtual photons:

\begin{equation}
\gamma^{*} d \rightarrow p p X \ .
\end{equation}

In the forward scattering approximation a virtual photon can be 
replaced by a real photon. Assuming that only a single $\pi^{-}$ were 
produced, the photon energy can be calculated from energy and 
momentum balance using momenta of the both protons, defined in 
the experiment. One can say that experimental distributions 
essentially exceed the  theoretical estimations made in the one 
particle frameworks. The background event contribution that it 
does not take into account could be a source of this difference.
Since there is no additional equation, one cannot distinguish the 
background events with one and more neutral particles. Although 
an admixture of the background process with additional $\pi^{0}$ 
according to the experimental estimations is small, we decided to 
calculate it directly. It should be noted that the maximum value 
of the missing mass in our experiment\cite{sid}:
\begin{equation}
 M_{mis}=\sqrt{(M_{d}+E_{e}-E_{1}-E_{2})^{2}-(\vec{p}_{e}-
\vec{p}_{1}-\vec{p}_{2})^{2}}
\end{equation}
is obviosly less than the $ \rho $ -- meson mass, so background 
from $ \rho $-- production is absent.

\begin{figure}[ht]
\centering \includegraphics[width=0.50\textwidth]{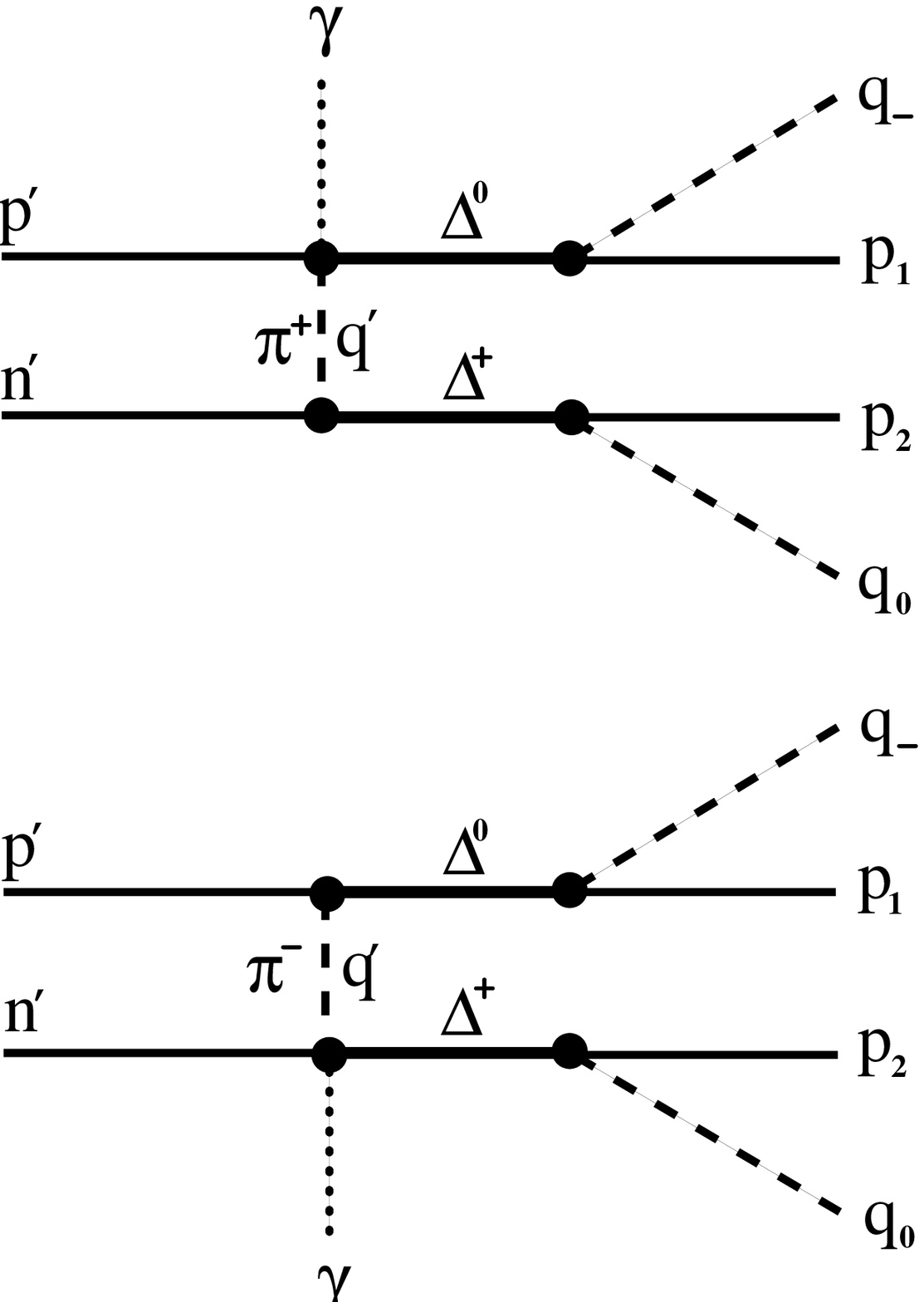}
\caption{} \label{f1}
\end{figure}

We have studied the background process $\gamma d \rightarrow p 
p\pi^{-}\pi^{0}$ as two different possibilities. The first is 
the production of the  $\pi^{-} \pi^{0}$ on the neutron -- let 
us call it the pole contribution and the two particle process 
of the $\Delta^{+} \ \Delta^{0}$  production when the final  
$\pi p$ systems comes from the decay of $\Delta$ according to 
the diagrams depicted in Fig.\ref{f1}. These channels do not exhaust 
all possibilities. The model of the production of two pions  
on the nucleon and the  deuteron was developed in the works 
\cite{gomes1, gomes2}. We used it for the  tensor polarized 
deuteron and large momenta of the decayed protons in our 
isotopic channel.

The marix elements of the pole and double $\Delta$  reaction 
mechanisms are:
\begin{eqnarray}
iM^{(pole)} \phantom{}^{m_{1} \ m_{2}}_{M_{d} \ \lambda } = 
\phantom{iiiiiiiiiiiiiiiii} & & \nonumber \\
\frac{\sqrt{2}}{3}e \left(\frac{f^{*}}{\mu}\right)^{2} 
\sum_{m'_{1}} \chi^{\dagger}_{m_{1}} \left( \frac{2}{3} 
\vec{q}_{-} \cdot \vec{\epsilon}_{\lambda} -\frac{i}{3}
\vec{q}_{-} 
\cdot [\vec{\epsilon}_{\lambda} \  \vec{\sigma}] \right) 
\chi_{m'_{1}}\times & & \nonumber \\
\times  G_{\Delta}(p_{1}+q_{0}) \Psi_{m'_{1} m_{2} ,M_{d}} 
(\vec{p_{2}}) - 
\phantom{iiiiiiiiiiii} & & \nonumber \\
-\frac{\sqrt{2}}{3}e \left(\frac{f^{*}}{\mu}\right)^{2} 
\sum_{m'_{1}} \chi^{\dagger}_{m_{2}} \left( \frac{2}{3} 
\vec{q}_{-} \cdot \vec{\epsilon}_{\lambda} -\frac{i}{3}
\vec{q}_{-} 
\cdot [\vec{\epsilon}_{\lambda} \  \vec{\sigma}] \right) 
\chi_{m'_{1}}\times & & \nonumber \\ 
\times G_{\Delta}(p_{2}+q_{0})\Psi_{m'_{1} m_{1} ,M_{d}} 
(\vec{p_{1}})\phantom{iiiiiiiiiiiii} & & 
\end{eqnarray}

\begin{eqnarray}
iM^{(delta)}\phantom{}^{m_{1} \ m_{2}}_{M_{d} \ \lambda } = 
\phantom{iiiiiiiilllllllliiiiiiiiiii} & & \nonumber \\
i \frac{\sqrt{2}}{9} e \left(\frac{f^{*}}{\mu}\right)^{4}
\int \frac{d^{3}n'}{(2 \pi)^{3}}\sum_{m'_{1} \ m'_{2}} 
\ \chi^{\dagger}_{m_{1}} \left( \frac{2}{3} \vec{q}_{-} 
\cdot \vec{\epsilon}_{\lambda} -\frac{i}{3}\vec{q}_{-} 
\cdot [\vec{\epsilon}_{\lambda} \  \vec{\sigma}] \right) 
\chi_{m'_{1}} \ \times
& & \nonumber \\
\times \chi^{\dagger}_{m_{2}} \left(\frac{2}{3}
\vec{q}_{0} \cdot \vec{q'} - \frac{i}{3} \cdot [\vec{q}_{0} 
\  \vec{\sigma}] \right) \chi_{m'_{2}}
F^{2}(q'^{2}) G_{\Delta}(p_{1}+q_{-})\times 
\phantom{iiiiiii} & & \nonumber \\
\times G_{\Delta}(p_{2}+q_{0}) \  \frac{1}{q'^{2}-\mu^{2}+i 
\epsilon} 
\Psi_{m'_{1} m'_{2} ,M_{d}} (\vec{n'}) + \phantom{iiiiiiiiiiiiii}  
& & \nonumber \\
 + i\frac{\sqrt{2}}{9} e \left(\frac{f^{*}}{\mu}\right)^{4} 
\int \frac{d^{3}p'}{(2 \pi)^{3}}\sum_{m'_{1} \ m'_{2}} 
\ \chi^{\dagger}_{m_{1}} \left(\frac{2}{3}
\vec{q}_{-} \cdot \vec{q'} - \frac{i}{3} \cdot [\vec{q}_{-} 
\  \vec{\sigma}] \right) \chi_{m'_{1}} \ \times
& & \nonumber \\
\times \chi^{\dagger}_{m_{2}} \left( \frac{2}{3} \vec{q}_{0} 
\cdot \vec{\epsilon}_{\lambda} -\frac{i}{3}\vec{q}_{0} 
\cdot [\vec{\epsilon}_{\lambda} \  \vec{\sigma}] \right) 
\chi_{m'_{2}}
F^{2}(q'^{2})G_{\Delta}(p_{1}+q_{-}) \times \phantom{ii}  
& & \nonumber \\
\times G_{\Delta}(p_{2}+q_{0}) \  \frac{1}{q'^{2}-\mu^{2}+i 
\epsilon} 
 \Psi_{m'_{1} m'_{2} ,M_{d}} (\vec{p'}) - \phantom{iiiiiiiiiiiiii}  
& & \nonumber 
\end{eqnarray}

\begin{eqnarray}
- (1 \leftrightarrow 2 ) \ . \phantom{i}   & &
\end{eqnarray}
where $M_{d}, \ \lambda, \  m_{1}, \  m_{2}$  are spin projection of 
the initial and final particles; $e$ is the electron charge; $\mu, 
\  M_{\Delta}$ are the meson and $\Delta$ masses;
$p_{1}, \  p_{2}, \  n', \  p', \ q_{0}, \  q_{-}$ are four momenta of 
the nucleons and pions according to diagrams depicted in Fig.1. 
Coupling constant $f^{*}/4\pi=0.36$. $\epsilon_{\lambda}$ is the photon 
polarization vector (in Coulomb gauge 
$\epsilon_{0}=0, \ \vec{\epsilon}\cdot\vec{k}=0 $),\  $F(q^{2})$  
a monopole form factor with $\Lambda=1.3 GeV $, \ 
$\Psi_{m_{1} m_{2} ,M_{d}}(\vec{p})$  is the deuteron relative wave 
function in momentum space. For the $\Delta$ propagator we take the formula:
\begin{equation}
G_{\Delta}(p)=\frac{1}{\sqrt{s}-m_{\Delta}+\frac{i}{2}\Gamma_{\Delta}(s)}
\frac{M_{\Delta}}{E_{\Delta}(\vec{p})} \ .
\end{equation}
The two proton yield from the reaction $e d \rightarrow p p \pi^{-} 
\pi^{0} e'$:
\begin{equation}
Y=L \int f(\omega) \frac{d^{8} \sigma_{\gamma^{*}}(\omega)}{d^{3}
p_{1}d^{3}p_{2}d\Omega_{q}} \ d^{3}p_{1}d^{3}p_{2}  d\Omega_{q} \ ,
\end{equation}
where $L$-experimental luminosity, $\Omega_{q}$ is solid angle of the 
relative momentum of the negative and neutral pions.
We have selected ($\theta_{i},\ \phi_{i},\ p_{i} $) of the protons from 
our experimental regions \cite{sid} and integrated the differential cross 
section of the background process using the Monte Carlo method. 

\begin{figure}[ht]
\centering \includegraphics[width=0.55\textwidth]{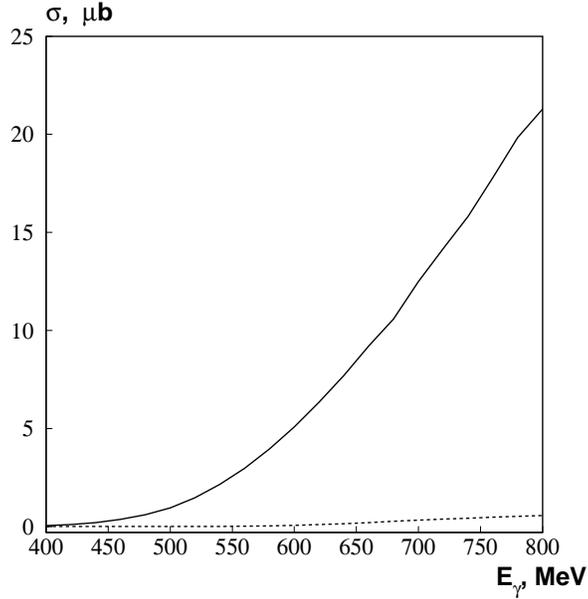}
\parbox[t]{0.70\textwidth}{
\caption{The total cross section of the $\pi^- \pi^0$ 
photoproduction on the deuteron target as a function of the 
photon energy (solid line - pole mechanism, 
dashed line - $\Delta^+ \Delta^0$  production).} 
\label{f2}
}
\end{figure}
 
Both pions can fly in $4 \pi$ geometry, so we have itegrated differential 
cross section by  $d\Omega_{q}$ over the full solid angle. The pion 
energies was determined by the conservation laws.
The energy spectrum of the virtual photons has the form presented 
in \cite{dalitz}.

\begin{figure}
\includegraphics[width=0.47\textwidth]{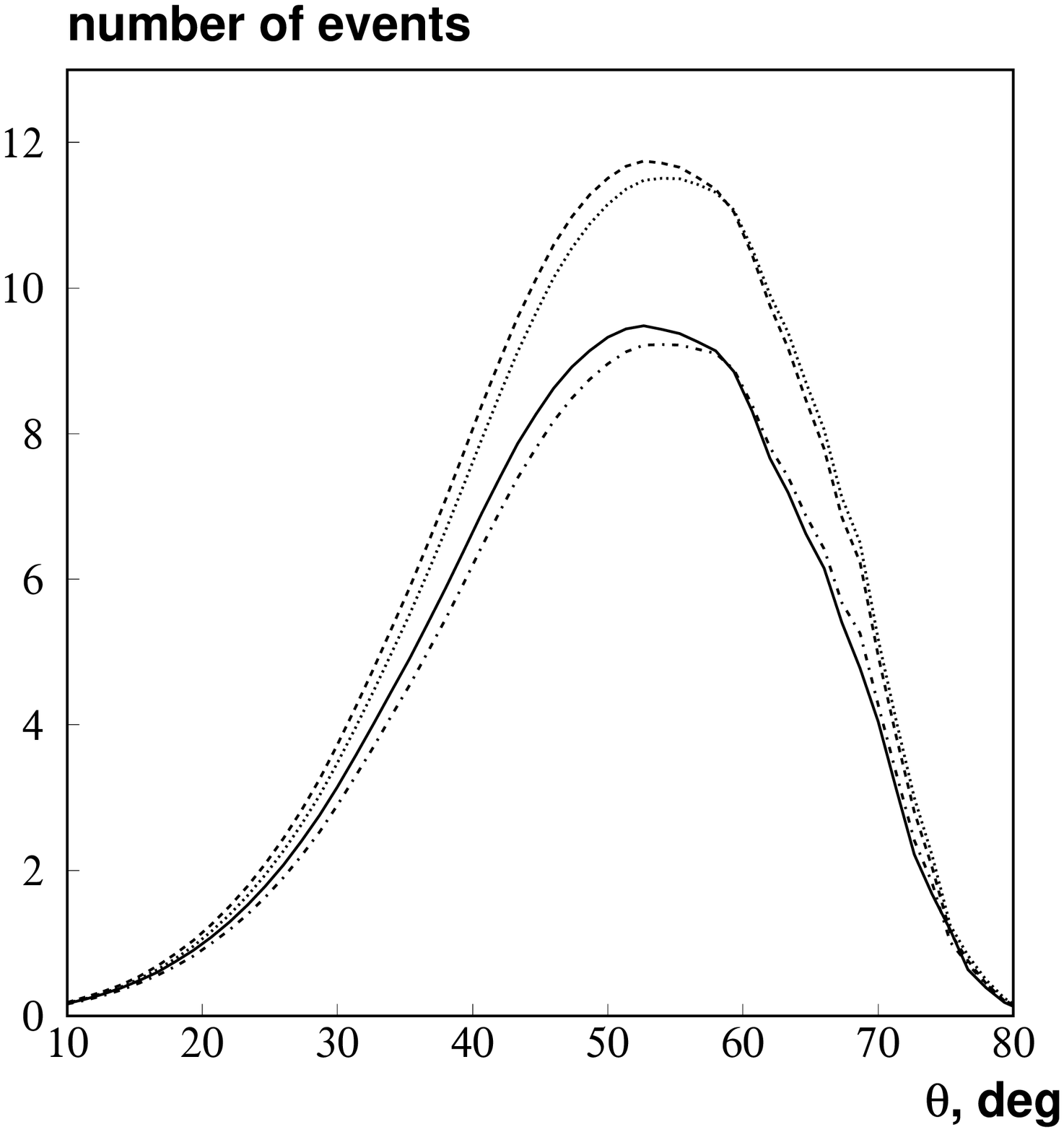}
\hfill
\includegraphics[width=0.47\textwidth]{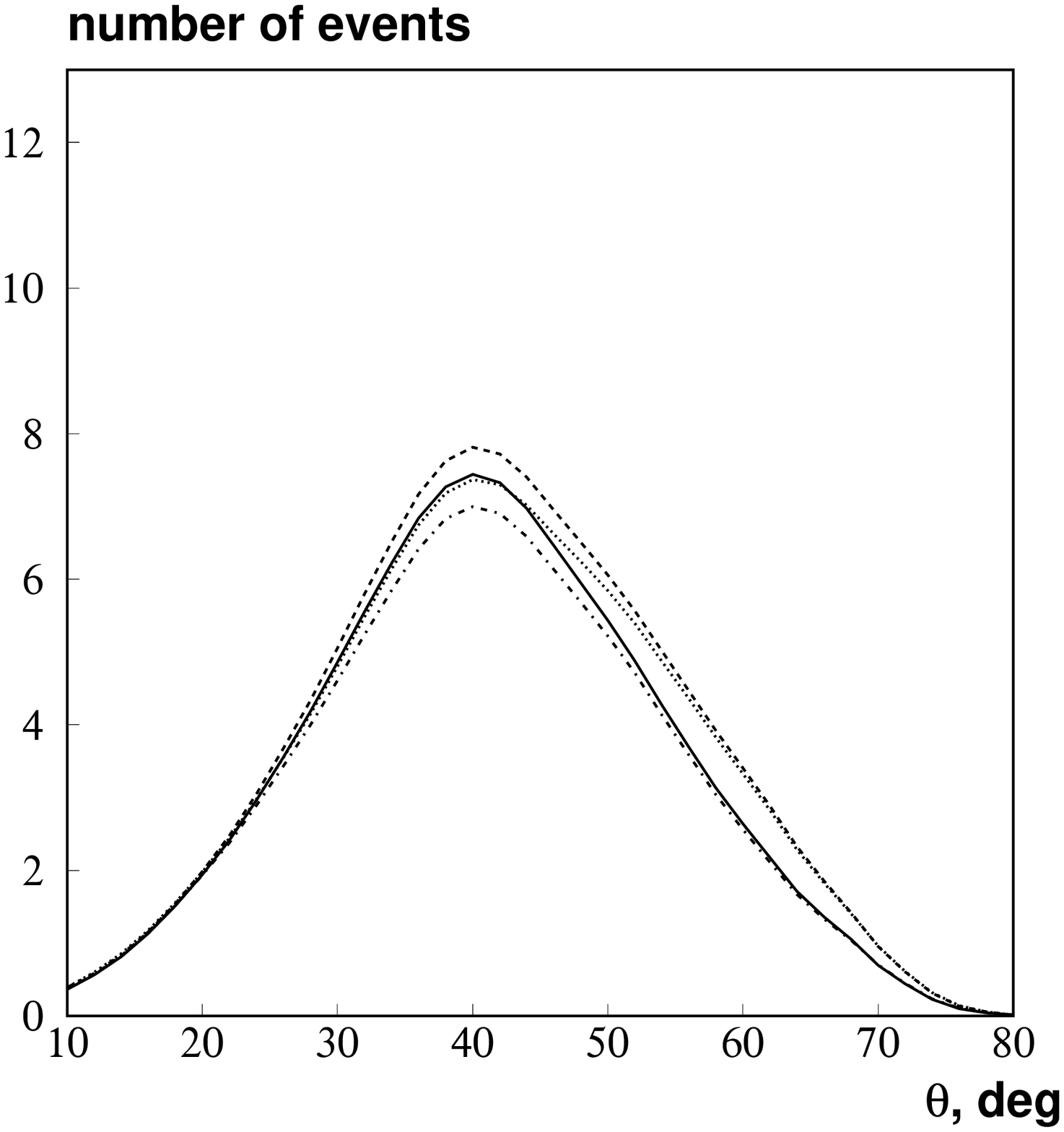}
\parbox[t]{0.47\textwidth}{\caption{
The yield of the $\pi^-\pi^0$ production the pole mechanism
as a function of the angle between proton detector axis and photon
beam direction.}
\label{f3}
}
\hfill
\parbox[t]{0.47\textwidth}{\caption{
The yield of the $\pi^-\pi^0$ production for double 
$\Delta$ mechanism as a function of the angle between proton detector 
axis and photon beam direction.}
\label{f4}
}
\end{figure}

For the checking  we estimated the total cross section of the 
$\pi^{-} \pi^{0}$ photoproduction on the deuteron target by the same method 
for both mechanisms and presented them in Fig.\ref{f2}. The total cross section
as a function of the photon energy for the studied isotopic channel agrees 
with the results of \cite{gomes2} for another isotopic channel if the 
$\Delta^{+}, \Delta^{0}$ decay would be considered.

The results of the estimation of $Y$  for the two reaction mechanisms are 
presented in Fig.\ref{f3} and Fig.\ref{f4} for two signes of the deuteron target tensor 
polarization: ($Pzz = 0.573$ --  solid and dashed lines, 
\ $Pzz= -0.573 $ -- dotted and dash-dotted lines) and for two directions 
of the magnetic field: (solid and dotted lines -- $\vec{H}$ in the plane 
of the proton detector axes, dashed and dash-dotted lines -- $\vec{H}$  
perpendicular to that plane).

These four curves is shown only in illustrative 
purposes as at our experimental conditions they cannot be distinguished.
These background events are presented as a function of the  angle between 
the proton detector axis and photon beam direction when the proton detectors 
are placed symmetrically about this direction.
One can see that the total cross section, calculated with pole  matrix 
elements, drastically exceeds that for the second mechanism.
In order to estimate the contributios of both mechanisms
into the calculated yield at the experimental region \cite{sid},
it is necessary to put the angle $\theta$ in Figs.\ref{f3}, \ref{f4} 
equal to $75^{\circ}$.
So the background yield amounts to about three events. It is
too small to be represnted it together with experimental and
theoretical distributions of the proton-proton events from the 
main process \cite{sid}. Indeed, the whole experimental yield is about
440 events at one system of the proton-proton registration
and the theoretical yield, calculated in a spectator model,
is about 250 events.

 A assumption  about the appreciable contribution of these background events  
turned out to be unfounded. A number of theoretical models were used for explanation  of experimental data \cite{sid}.
The approaches based on the traditional concepts seem to be unable to reproduce all features of the 
proton-proton production from the deuteron \cite{sid, few}.


\end{document}